\documentstyle[12pt]{article}
\font\tenrm=cmr10
\begin{document}
\baselineskip 24pt
\newcommand{\be}{\begin{equation}}
\newcommand{\ee}{\end{equation}}
\newcommand{\sheptitle}
{Squark Contributions to Higgs Boson Masses \\
in the Next--to--Minimal Supersymmetric \\
Standard Model}

\newcommand{\shepauthor}
{T. Elliott, S. F. King and P. L. White}

\newcommand{\shepaddress}
{Physics Department\\University of Southampton\\Southampton
        \\SO9 5NH\\U.K.}

\newcommand{\shepabstract}
{Within the context of an effective potential formalism we calculate
the contribution to Higgs boson masses in the next--to--minimal
supersymmetric standard model from squark loops. We then supplement
a previously performed renormalisation group analysis of the Higgs
sector of this model with these results in order to determine the
shift in the bound on the lightest CP-even Higgs boson mass as a
result of squark effects. The improved bound on the lightest neutral
CP-even Higgs boson mass, including squark contributions, is
$m_h \leq 146 \ \ (139, 149)$ GeV for $m_t = 90 \ \ (140, 190)$ GeV.
For $m_t = 190$ GeV squark effects contribute 23 GeV to the bound,
with smaller contributions for small $m_t$ values.}

\begin{titlepage}
\begin{flushright}
SHEP 92/93-18 \\ hep-ph/9305282
\end{flushright}
\vspace{.4in}
\begin{center}
{\Large{\bf \sheptitle}}
\bigskip \\ \shepauthor \\ {\it \shepaddress} \\ \vspace{.5in}
{\bf Abstract} \bigskip \end{center} \setcounter{page}{0}
\shepabstract
\end{titlepage}

\section*{}

The next--to--minimal supersymmetric standard model \cite{fayet,durand}
(NMSSM) provides an elegant extension of the minimal supersymmetric
standard model \cite{nilles} (MSSM) which eliminates the
$\mu$--problem, the appearance of a mass--scale in the superpotential.
The NMSSM differs from the MSSM by the presence of a gauge singlet
field whose vacuum expectation value (vev) plays the role of the mass
parameter $\mu$ in the MSSM, and is defined by  the superpotential
\be
W = h_t Q H_2 t^c + \lambda N H_1 H_2 - \frac{1}{3} k N^3 + \dots,
\label{superpotential}
\ee
where the superfield $Q^T=(t_L,b_L)$ contains the left--handed top and
bottom quarks, and $t^c$ contains the charge conjugate of the
right--handed top quark; $H_1$ and $H_2$ are the usual Higgs doublet
superfields and $N$ is the Higgs gauge singlet superfield; the ellipsis
represents terms whose relatively small couplings play no role in our
analysis (in particular, we assume that the bottom quark Yukawa coupling,
$h_b$, satisfies $h_b \ll h_t$).
The fields $H_1^T=(H_1^0, H_1^-)$, $H_2^T=(H_2^+, H_2^0)$ and
$N$ develop vevs which may be assumed to be of the form \cite{fayet}
\be
<H_1> = \left( \begin{array}{c} \nu_1 \\ 0 \end{array} \right), \ \ \
<H_2> = \left( \begin{array}{c} 0 \\ \nu_2 \end{array} \right), \ \ \
<N> = x,
\ee
where $\nu_1$, $\nu_2$ and $x$ are real, and $\sqrt{\nu_1^2+\nu_2^2}
= \nu =174$ GeV. The low energy physical spectrum of the Higgs scalars
consists of 3 CP-even neutral states, 2 CP-odd neutral states, and
2 charged scalars. A third CP-odd state is a Goldstone mode which
becomes
the longitudinal component of the $Z^0$, while a further two charged
scalars
become those of the $W^\pm$'s.

The tree--level bound on the lightest CP-even Higgs scalar mass in the
NMSSM \cite{durand},
\be
m_h^2 \leq M_Z^2 + (\lambda^2 \nu^2 - M_Z^2) \sin^2 2 \beta,
\label{blob}
\ee
where $\tan \beta = \nu_2 / \nu_1$, and radiative corrections to it
\cite{terveldhuis,ellwanger},
have been the subject of much discussion. In Ref. \cite{terveldhuis}
renormalisation group (RG) analyses were performed assuming either
supersymmetric (SUSY) RG equations down to near the top quark mass,
or the
existence of only one light Higgs boson with a mass below the SUSY
breaking scale. These analyses assumed degenerate squarks decoupling
below
their common mass scale $M_{SUSY}$. In Refs. \cite{ellwanger,ellwanger2},
within the effective potential formalism, squark
effects have been considered in the NMSSM. However, no systematic
attempt
was made to study their effects on the bound in Eq.~(\ref{blob}),
but rather one slice through parameter space was examined
\cite{ellwanger2}.
Moreover, radiative corrections to the charged scalar
mass--squared matrix
were not presented, though these do not change the bound.

Recently we performed a low energy
renormalisation group analysis of the Higgs sector of the NMSSM and
found that $m_h \leq 146$ GeV \cite{elliott}. This analysis considered
the more general case of all the Higgs bosons having masses below the
SUSY breaking scale ($M_{SUSY}$), and used non--SUSY RG equations below
$M_{SUSY}$. However, in that analysis the squarks were assumed to be
degenerate at $M_{SUSY} = 1$ TeV. Here we supplement this calculation
with a computation of the stop and, where necessary, sbottom
contributions
to the CP-even, CP-odd and charged mass-squared matrices, from which we
may determine modifications to the bound due to a general spectrum of
squarks. We then attempt systematically to examine the squark
corrected bound
by surveying as much of the parameter space as practicable.

Our previous analysis worked exclusively with $V_0(\mu)$, $\mu =150$ GeV
being approximately the bound on the lightest CP-even state.
 From the superpotential in Eq.~(\ref{superpotential}) we may obtain the
pure Higgs part of $V_0$, where we have renormalised at the
$\overline{MS}$
scale $\mu$:
\clearpage
\begin{eqnarray}
V_0(\mu)  & = & \frac{1}{2} {\lambda}_1(H_1^{\dagger} H_1)^2
              + \frac{1}{2} {\lambda}_2(H_2^{\dagger} H_2)^2
              + ({\lambda}_3 +{\lambda}_4)(H_1^{\dagger} H_1)
                    (H_2^{\dagger} H_2)
                \nonumber \\
          & - & {\lambda}_4|H_2^{\dagger} H_1|^2
          + {\lambda}_5|N|^2|H_1|^2 + {\lambda}_6|N|^2|H_2|^2
                 \nonumber \\
          & + & {\lambda}_7(N^{\ast 2} H_1H_2 + h.c.)
                + {\lambda}_8|N|^4 \nonumber \\
          & + &  m_1^2|H_1|^2 + m_2^2|H_2|^2 + m_3^2|N|^2 \nonumber \\
          & - &  m_4(H_1H_2N + h.c.)- \frac{1}{3} m_5(N^3 + h.c.),
\label{lowpotential}
\end{eqnarray}
where $\lambda_i = \lambda_i(\mu)$ and $m_i=m_i(\mu)$. At $M_{SUSY}$
the couplings $\lambda_i$ and $m_i$ are given by:
\[
{\lambda}_1 = {\lambda}_2=(g_2^2 + g_1^2)/4,\ \ \
{\lambda}_3=(g_2^2 - g_1^2)/4,
\]
\[
{\lambda}_4 = {\lambda}^2-g_2^2/2,\ \ \
{\lambda}_5={\lambda}_6={\lambda}^2, \ \ \ {\lambda}_7=
-{\lambda}k, \ \ \
{\lambda}_8=k^2,
\]
\be
m_1 = m_{H_1}, \ \ \ m_2  =  m_{H_2}, \ \ \ m_3  =  m_N, \ \ \
m_4 = {\lambda}A_{\lambda}, \ \ \ m_5=kA_k.
\label{bc}
\ee
$g_1$ and $g_2$ are the $U(1)_Y$ and $SU(2)_L$ gauge coupling constants,
respectively, and $m_{H_i}$, $m_N$, $A_{\lambda}$ and $A_k$ are soft SUSY
breaking parameters. We use the RG equations \cite{elliott} to allow to us
calculate the couplings renormalised at $\mu =150$ GeV given their values
at $M_{SUSY}$.

In order the calculate the shifts in the mass--squared matrices due to
squarks, we employ the full one--loop effective potential. Much of our
analysis of the squark spectrum is similar to that of the MSSM in
Ref. \cite{ellis}, whose notation we follow closely.
The one--loop effective potential is given by
\begin{eqnarray}
V_1(Q) & = & V_0(Q) + \Delta V_1(Q), \nonumber \\
\Delta V_1(Q) & = & \frac{1}{64\pi^2} \mbox{Str} {\cal M}^4 \left(
\ln \frac{{\cal M}^2}{Q^2} - \frac{3}{2} \right),
\end{eqnarray}
where $V_0(Q)$ is the tree--level potential at the arbitrary
$\overline{MS}$
scale $Q$, $\mbox{Str}$ denotes the usual supertrace, and ${\cal M}^2$
is the
field--dependent mass--squared matrix. We shall take $Q = M_{SUSY} = 1$ TeV.
\footnote{This value of $Q$ is of course totally arbitrary. The value of
1 TeV is chosen to be consistent with our previous analysis. The results
are independent of $Q$ to one loop order.}
$V_0$ is restricted to the pure Higgs part of the tree--level potential.
Since our RG approach has already included the logarithmic effects of top
quark and Higgs boson loops we exclude these particles from the supertrace.
The logarithmic contributions to $\Delta V_1 (Q)$ from the top quark and
Higgs bosons may be absorbed into $V_0(Q)$ where $Q = 1$ TeV, to yield
$V_0(\mu)$ where $\mu = 150$ GeV, as discussed above. It only remains to
calculate $\Delta V_1(Q)$, where $Q = 1$ TeV, involving squark
contributions.

To calculate the shifts in the Higgs mass matrices as a result of squark
effects we require the field--dependent mass--squared matrices of the
squarks. Taking the contribution to the soft SUSY breaking potential
involving squarks to be
\be
\Delta V_{soft} = h_t A_t (\tilde{Q} H_2 \tilde{t^c} + h.c.) +
                  m_Q^2 | \tilde{Q} |^2 +m_T^2 | \tilde{t^c} |^2 +
                  m_B^2 | \tilde{b^c} |^2,
\ee
where $b^c$ contains the charge conjugate of the right--handed bottom
quark, and tildes denote the scalar components of the superfields,
together with the superpotential in Eq.~(\ref{superpotential}), leads
to the
field--dependent squark mass--squared matrix (ignoring contributions
proportional to gauge couplings and $h_b$)
\be
{\cal M}^2 = \left(
\begin{array}{cccc}
m_Q^2 + h_t^2 |H_2^0|^2 & \lambda h_t N H_1^0 + h_t A_t \bar{H_2^0}
        & -h_t^2 \bar{H_2^0} H_2^+ & 0 \\
\lambda h_t \bar{N} \bar{H_1^0} + h_t A_t H_2^0 & m_T^2
       + h_t^2 (|H_2^0|^2+H_2^+H_2^-) & \lambda h_t \bar{N} H_1^+
       - h_t A_t H_2^+ & 0 \\
-h_t^2 H_2^0 H_2^- & \lambda h_t N H_1^-
       - h_t A_t H_2^- & m_Q^2 + h_t^2 H_2^+ H_2^- & 0 \\
0 & 0 & 0 & m_B^2
\end{array} \right),
\label{squarkmatrix}
\ee
in the basis
$\{ \tilde{t_L},\bar{\tilde{t_R^c}},\tilde{b_L},\bar{\tilde{b_R^c}} \}$,
where a bar denotes complex conjugation. Manifestly, one eigenvalue is
field--independent and may be discarded. Only the upper
$2 \times 2$ submatrix
contributes to the CP-even and CP-odd mass--squared matrices, since the
charged fields have zero vevs in order not to break QED, whereas the
upper $3 \times 3$ submatrix contributes to the charged mass--squared
matrix. By differentiating $\Delta V_1$ once with respect to the fields we
may obtain the shifts in the minimisation conditions induced by squark loops,
and twice will yield the shifts in the mass matrices.
\footnote{In fact, this is only approximately true due to Higgs
self--energy
corrections; these are expected to be small for the lightest Higgs bosons.}

In the basis $\{ H_1, H_2, N \}$ we have the following mass-squared
matrices, after ensuring that the full one--loop potential is correctly
minimised. The couplings $\lambda_i$ are those obtained from the potential
in Eq.~(\ref{lowpotential}) renormalised at the scale $\mu=150$ GeV.
For notational simplicity we drop the tildes. Moreover, we work in the
basis
of mass eigenstates $\{ t_1 , t_2 , b_1 , b_2 \}$ (though, since $h_b$, the
bottom quark Yukawa coupling, is taken to be zero, $b_2$ never contributes
to the mass--squared matrices), where the mass of the $t_1$ squark is
$m_{t_1}$, and so on for the other squarks. The CP-even (scalar)
mass--squared matrix is
\be
M^2_s = M^2 + \delta M^2,
\ee
where
\begin{eqnarray}
M^2 & = &
\left( \begin{array}{ccc}
2\lambda_1 \nu_1^2 & 2(\lambda_3+\lambda_4) \nu_1 \nu_2
   & 2\lambda_5 x \nu_1 \\
2(\lambda_3+\lambda_4) \nu_1 \nu_2 & 2\lambda_2 \nu_2^2
   & 2\lambda_6 x \nu_2 \\
2\lambda_5 x \nu_1 & 2\lambda_6 x \nu_2 & 4\lambda_8 x^2 - m_5 x
\end{array} \right) \nonumber \\
 & + & \left( \begin{array}{ccc}
\tan \beta [m_4x-\lambda_7x^2] & - [m_4x-\lambda_7x^2]
    & -\frac{\nu_2}{x} [m_4x-2\lambda_7x^2] \\
- [m_4x-\lambda_7x^2] & \cot \beta [m_4x-\lambda_7x^2]
    & -\frac{\nu_1}{x} [m_4x-2\lambda_7x^2] \\
-\frac{\nu_2}{x} [m_4x-2\lambda_7x^2]
   & -\frac{\nu_1}{x} [m_4x-2\lambda_7x^2] & \frac{\nu_1 \nu_2}{x^2} [m_4x]
\end{array} \right)
\end{eqnarray}
and
\be
\delta M^2 =
\left( \begin{array}{ccc}
\Delta_{11}^2 & \Delta_{12}^2 & \Delta_{13}^2 \\
\Delta_{12}^2 & \Delta_{22}^2 & \Delta_{23}^2 \\
\Delta_{13}^2 & \Delta_{23}^2 & \Delta_{33}^2
\end{array} \right)
+
\left( \begin{array}{ccc}
\tan \beta & -1 & -\frac{\nu_2}{x} \\
-1 & \cot \beta & -\frac{\nu_1}{x} \\
-\frac{\nu_2}{x} & -\frac{\nu_1}{x} & \frac{\nu_1 \nu_2}{x^2}
\end{array} \right)
\Delta_p^2.
\label{blob1}
\ee
The $\Delta_{ij}^2$ and $\Delta_p^2$ are given by
\begin{eqnarray}
\Delta_p^2    & = & \frac{3}{16\pi^2} h_t^2.(\lambda x).A_t.
                    f(m_{t_1}^2,m_{t_2}^2), \nonumber \\
\Delta_{11}^2 & = & \frac{3}{8\pi^2} h_t^4 \nu_2^2.(\lambda x)^2.
                    \left(
                    \frac{A_t+\lambda x \cot \beta}{m_{t_2}^2-m_{t_1}^2}
                    \right)^2
                    g(m_{t_1}^2,m_{t_2}^2), \nonumber \\
\Delta_{22}^2 & = & \frac{3}{8\pi^2} h_t^4 \nu_2^2.\left(
                    \ln \frac{m_{t_1}^2 m_{t_2}^2}{M_{SUSY}^4} +
                    \frac{2A_t(A_t+\lambda x \cot \beta)}
                         {m_{t_2}^2-m_{t_1}^2}
                    \ln \frac{m_{t_2}^2}{m_{t_1}^2} \right) \nonumber \\
              & + & \frac{3}{8\pi^2} h_t^4 \nu_2^2.
                    \left(
                    \frac{A_t(A_t+\lambda x \cot \beta)}
                         {m_{t_2}^2-m_{t_1}^2}
                    \right)^2
                    g(m_{t_1}^2,m_{t_2}^2), \nonumber \\
\Delta_{33}^2 & = & \frac{3}{8\pi^2} h_t^4 \nu_2^2.(\lambda \nu_1)^2.
                    \left( \frac{A_t+\lambda x \cot \beta}
                                {m_{t_2}^2-m_{t_1}^2} \right)^2
                    g(m_{t_1}^2,m_{t_2}^2), \nonumber \\
\Delta_{12}^2 & = & \frac{3}{8\pi^2} h_t^4 \nu_2^2.(\lambda x).
                    \left(
                    \frac{A_t+\lambda x \cot \beta}
                         {m_{t_2}^2-m_{t_1}^2}
                    \right)
                    \left( \ln \frac{m_{t_2}^2}{m_{t_1}^2} +
                    \frac{A_t(A_t+\lambda x \cot \beta)}
                         {m_{t_2}^2-m_{t_1}^2}
                    g(m_{t_1}^2,m_{t_2}^2) \right), \nonumber \\
\Delta_{13}^2 & = & \frac{3}{8\pi^2} h_t^4
         \nu_2^2.(\lambda x).(\lambda \nu_1).
                    \left(
                    \frac{A_t+\lambda x \cot \beta}
                         {m_{t_2}^2-m_{t_1}^2}
                    \right)^2
                    g(m_{t_1}^2,m_{t_2}^2) \nonumber \\
              & - & \frac{3}{8\pi^2} h_t^2.(\lambda x).(\lambda \nu_1).
                    f(m_{t_1}^2,m_{t_2}^2), \nonumber \\
\Delta_{23}^2 & = & \frac{\nu_1}{x} \Delta_{12}^2,
\label{blob2}
\end{eqnarray}
where the functions $f$ and $g$ are defined by
\begin{eqnarray}
f(m_{t_1}^2,m_{t_2}^2) & = & \frac{1}{m_{t_2}^2-m_{t_1}^2} \left[
                             m_{t_1}^2 \ln \frac{m_{t_1}^2}{M_{SUSY}^2}
        - m_{t_1}^2
                           - m_{t_2}^2 \ln \frac{m_{t_2}^2}{M_{SUSY}^2}
        + m_{t_2}^2
                             \right], \nonumber \\
g(m_{t_1}^2,m_{t_2}^2) & = & \frac{1}{m_{t_1}^2-m_{t_2}^2} \left[
                           (m_{t_1}^2+m_{t_2}^2) \ln \frac{m_{t_2}^2}
                                     {m_{t_1}^2}
                           +2(m_{t_1}^2-m_{t_2}^2) \right] .
\end{eqnarray}
If the squarks are degenerate in mass, then equation
Eq.~(\ref{squarkmatrix})
implies that $A_t+\lambda x \cot \beta =0$. Furthermore, if
$m_{t_1} = m_{t_2} = M_{SUSY}$,
then Eq.~(\ref{blob1}) and Eq.~(\ref{blob2}) imply that the squark
contribution
to the CP-even mass--squared matrix vanishes. This is the limit of our
previous analysis.

The CP-odd (pseudoscalar) mass--squared matrix is
\be
M^2_p = \tilde{M}^2 + \delta \tilde{M}^2,
\ee
where
\be
\tilde{M}^2 =
\left( \begin{array}{ccc}
\tan \beta [m_4x-\lambda_7x^2] & [m_4x-\lambda_7x^2]
     & \frac{\nu_2}{x} [m_4x+2\lambda_7x^2] \\
\mbox{} [m_4x-\lambda_7x^2] & \cot \beta [m_4x-\lambda_7x^2]
     & \frac{\nu_1}{x} [m_4x+2\lambda_7x^2] \\
\frac{\nu_2}{x} [m_4x+2\lambda_7x^2]
     & \frac{\nu_1}{x} [m_4x+2\lambda_7x^2]
     & 3m_5x + \frac{\nu_1 \nu_2}{x^2} [m_4x-4\lambda_7x^2]
\end{array} \right)
\ee
and
\be
\delta \tilde{M}^2 =
\left( \begin{array}{ccc}
\tan \beta & 1 & \frac{\nu_2}{x} \\
1 & \cot \beta & \frac{\nu_1}{x} \\
\frac{\nu_2}{x} & \frac{\nu_1}{x} & \frac{\nu_1 \nu_2}{x^2}
\end{array} \right) \Delta_p^2.
\ee
Finally, the charged mass--squared matrix is
\be
M_c^2 =
\left( \begin{array}{cc}
\tan \beta & 1 \\
1 & \cot \beta
\end{array} \right)
(m_4x-\lambda_7x^2-\lambda_4 \nu_1 \nu_2 + \Delta_c^2),
\label{chargematrix}
\ee
where
\be
\Delta_c^2 = \frac{3}{16\pi^2} \sum_{m_a \in \{ m_{t_1}, m_{t_2},
    m_{b_1} \} }
m_a^2 \left( \ln \frac{m_a^2}{M_{SUSY}^2} - 1 \right)
\frac{\partial^2 m_a^2}{\partial H_1^- \partial H_2^+} \left|_{vevs}
         \right. ,
\ee
and
\begin{eqnarray}
\frac{\partial^2 m_{t_1}^2}{\partial H_1^- \partial H_2^+}
\left|_{vevs} \right. & = &
-\frac{h_t^4 \nu_2^2.(\lambda x)^2. \cot \beta}
      {(m_{t_1}^2-m_{t_2}^2)(m_{t_1}^2-m_{b_1}^2)}
-\frac{h_t^2.(\lambda x).A_t}{m_{t_1}^2-m_{t_2}^2}, \nonumber \\
\frac{\partial^2 m_{t_2}^2}{\partial H_1^- \partial H_2^+}
\left|_{vevs} \right. & = &
-\frac{h_t^4 \nu_2^2.(\lambda x)^2. \cot \beta}
      {(m_{t_2}^2-m_{b_1}^2)(m_{t_2}^2-m_{t_1}^2)}
+\frac{h_t^2.(\lambda x).A_t}{m_{t_1}^2-m_{t_2}^2}, \nonumber \\
\frac{\partial^2 m_{b_1}^2}{\partial H_1^- \partial H_2^+}
\left|_{vevs} \right. & = &
-\frac{h_t^4 \nu_2^2.(\lambda x)^2. \cot \beta}
      {(m_{b_1}^2-m_{t_1}^2)(m_{b_1}^2-m_{t_2}^2)}.
\end{eqnarray}

The bound on the lightest CP-even Higgs mass is a consequence of the fact
that the minimum eigenvalue of $M_s^2$ is bounded by the minimum
eigenvalue
of the upper $2 \times 2$ submatrix of $M_s^2$. Using
Eq.~(\ref{chargematrix})
to obtain the physical charged Higgs mass squared, $m_c^2$, in terms of
$m_4$,
\be
m_4x-\lambda_7x^2 = \frac{1}{2} (m_c^2+\lambda_4 \nu^2)
              \sin 2\beta - \Delta_c^2,
\ee
we may eliminate $m_4$ from $M_s^2$ in favour of $m_c^2$, and thus
write
the upper $2 \times 2$ submatrix in the form
\be
M_s'^2 = M'^2 + \delta M'^2,
\ee
where
\be
M'^2 = \left( \begin{array}{cc}
2\lambda_1\nu_1^2 & 2(\lambda_3+\lambda_4)\nu_1\nu_2 \\
2(\lambda_3+\lambda_4)\nu_1\nu_2 & 2\lambda_2\nu_2^2
\end{array} \right)
+
\left( \begin{array}{cc}
\tan \beta & -1 \\
-1 & \cot \beta
\end{array} \right)
\frac{1}{2}(m_c^2+\lambda_4\nu^2)\sin 2\beta,
\label{treematrix}
\ee
and
\be
\delta M'^2 = \left( \begin{array}{cc}
\Delta_{11}^2 & \Delta_{12}^2 \\
\Delta_{12}^2 & \Delta_{22}^2
\end{array} \right)
+
\left( \begin{array}{cc}
\tan \beta & -1 \\
-1 & \cot \beta
\end{array} \right)
(\Delta_p^2 - \Delta_c^2).
\label{reduceparameters}
\ee
It is clear from the form of the second term on the
right--hand--side of
Eq.~(\ref{reduceparameters}) that the factor $\Delta_p^2 - \Delta_c^2$
does not change the bound at all, but merely serves, for a fixed
   $\tan \beta$,
to shift the bound, when plotted as a function of $m_c$, to the left
   or right.
We may then drop this term, since its presence may be re-parametrised by
a shift in the free parameter $m_c$. In this way $m_{b_1}$ is
eliminated.

 From Eq.~(\ref{treematrix}) and Eq.~(\ref{reduceparameters}) it is a
  simple
matter to determine the shift in the minimum eigenvalue
of $M'^2$ due to squark effects. Defining $A = (M'^2)_{11}$,
$B = (M'^2)_{12}$ and $C = (M'^2)_{22}$, the shift is given by
\be
\frac{1}{2} \left[ \Delta^2_{11} + \Delta^2_{22} -
\frac{(A - C)(\Delta^2_{11} - \Delta^2_{22}) + 4 B \Delta^2_{12}}
     {\sqrt{(A - C)^2 + 4 B^2}} \right].
\label{squarkshift}
\ee
Because this shift is a one loop effect, the couplings $\lambda_i$ in $A$,
$B$ and $C$ may be evaluated at any renormalisation point, since the
difference between a coupling evaluated at two different renormalisation
points is also a one loop effect, thus giving an overall error at the
two loop level, which we neglect in our approximation.

In our previous analysis \cite{elliott} we used triviality limits on the
couplings $h_t$, $\lambda$ and $k$ \cite{binetruy}
to determine the bound on the lightest CP-even
Higgs boson mass $m_h^0$ not including general squark effects, that is,
with
$\delta M'^2 \equiv 0$. Let the triviality limit on $\lambda$ for a given
$h_t$ be $\lambda_{max}$. In table \ref{table} we reproduce the
values of $h_t$ and $\lambda_{max}$ which generated the bound $m_h^0$
of our previous analysis --- all other values, for a
given top quark mass, resulted in smaller values of the bound. It
 transpires that $k = 0$ in all cases.
\footnote{Taking $k=0$ gives rise to an axion in the physical spectrum.
However, a small, non--zero value of $k$ is sufficient to give the
would--be axion a mass of several tens of GeV. Such a value of $k$ does
not
invalidate our calculations.}
In order to maximise the bound resulting from
Eq.~(\ref{treematrix}) we take the limit $m_c = \infty$, thus eliminating
$m_c$ as a parameter. As a function of $m_c$ the bound approaches its
maximum asymptotically as $m_c \rightarrow \infty$. The approach is rapid,
with $m_c \sim 200$ GeV being a good approximation to $m_c = \infty$.
Thus, the bound resulting from the purely formal procedure of taking the
limit $m_c = \infty$ is not unphysical \cite{elliott}. Had this not been
the
case, then certainly we would have an upper bound, but one perhaps not
capable of realisation in an actual spectrum.

\begin{table}
\caption{\tenrm \baselineskip=12pt
Lightest CP-even Higgs mass bound in the NMSSM including general
squark
effects. In row 1 is the top mass, $m_t$, in GeV; in rows 2 and 3
those
values of $h_t(M_{SUSY})$ and ${\lambda}_{max}(M_{SUSY})$,
which produce the bound in our previous analysis, this being in row 4,
in
GeV; in row 5 is the contribution to the bound resulting from a general
squark
mass spectrum rather than degenerate squarks with mass $M_{SUSY}$,
in GeV;
in row 6 is the lightest CP-even Higgs mass bound including general
squark
effects, in GeV; in rows 7 and 8 are
those value of $m_{t_1}$ and $A_t$, in GeV, which generate the bound in
row 6; $m_{t_2} = 1$ TeV.}
\begin{center}
\begin{tabular}{|c|r|r|r|r|r|r|r|r|r|r|r|}
\hline
$m_t$ & 90 & 100 & 110 & 120 & 130 & 140 & 150 & 160 & 170
& 180 & 190 \\
\hline
$h_t$ & 0.60 & 0.67 & 0.73 & 0.79 & 0.84 & 0.88 & 0.92 & 0.95
           & 0.98 & 1.00 & 1.03 \\
\hline
${\lambda}_{max}$ & 0.87 & 0.85 & 0.83 & 0.81 & 0.79 & 0.77 & 0.74
           & 0.71 & 0.67 & 0.63 & 0.50 \\
\hline
$m_h^0$  & 145 & 143 & 140 & 137 & 135 & 132 & 129 & 126 & 124
    & 123 & 126 \\
\hline
$\delta m_h$ &   1 &   2 &   3 &   4 &   6 &   8 &  10 &  13
                             &  16 &  20 &  23  \\
\hline
$m_h$   & 146 & 145 & 143 & 141 & 140 & 139 & 139 & 139 & 140
   & 143 & 149 \\
\hline
$m_{t_1}$ & 800 & 770 & 750 & 720 & 700 & 670 & 650 & 630 & 610
   & 590 & 570 \\
\hline
$A_t$ & 940 & 1000 & 1040 & 1130 & 1160 & 1270
                 & 1330 & 1410 & 1500 & 1630 & 1820 \\
\hline
\end{tabular}
\end{center}
\label{table}
\end{table}

We use the values of $h_t$ and $\lambda_{max}$ in table \ref{table} to
calculate the shift in the bound on the lightest CP-even Higgs
boson mass--squared given by Eq.~(\ref{squarkshift}). The shift in the
bound on the mass is denoted by $\delta m_h$. This is a hideously
complicated function of many parameters, including the squark masses
$m_{t_1}$ and $m_{t_2}$, the soft SUSY breaking parameter $A_t$,
the vev of the gauge singlet field $x$, and $m_c$. As a function
of $m_c$ the shift appears, numerically, to be maximised for $m_c$ small,
typically between 100 GeV and 200 GeV, with the difference between the
shifts
at these two points being less than 1 GeV. Thus, we set $m_c = 200$
GeV in the
general squark contributions to the bound. Doing this enables us to retain
the bound from our previous analysis derived
from Eq.~(\ref{treematrix}) in the limit $m_c = \infty$, and simply
maximise
the squark contributions separately and add them to our previous bound
to yield the new bound $m_h = m_h^0 + \delta m_h$.
The $x$ dependence is not too strong, with the ratio
$r = x / \nu$ typically taking values close to 10 in order to maximise
the
bound. Where this is not the case, the difference between the bound at
its
maximum, as a function of $x$, and that at $r=10$ is less than 1 GeV.
Thus, we set $r=10$, this giving a reliable indication of the maximum
shift.
We calculate the shift for various ranges of squark masses and a range of
values of $A_t$, and record the maximum value obtained. The squark masses
are allowed to vary between 20 GeV and 1000 GeV in steps of 10 GeV
(1000 GeV
is the upper limit since we take $M_{SUSY} = 1000$ GeV.) Given this
restriction, the value of $A_t$ which maximises the general squark
contribution to the bound never exceeds 2 TeV. We impose the constraint
$ 2h_t \nu_2 (A_t + \lambda x \cot \beta) \leq |m_{t_1}^2-m_{t_2}^2|$
which
follows from the form of the squark mass matrix in
Eq.~(\ref{squarkmatrix}).

In table \ref{table} and figure 1 we show the bound on the lightest
CP-even
Higgs mass $m_h$ including general squark effects (and $m_h^0$ for
comparison). Table \ref{table} also shows the values of $m_{t_1}$ and
$A_t$ used to generate the bound.
As $m_t$ increases, $m_{t_1}$ monotonically decreases, while $A_t$
monotonically increases. The other squark mass, $m_{t_2}$, takes the value
1 TeV over the whole range of top quark masses --- that one squark
mass takes
on its maximum permitted value to maximise the general squark
contribution to
the bound can be seen analytically. It will be noticed that these squark
masses and $A_t$ are somewhat larger than typically expected from Grand
Unification scenarios. However, we feel it wise not to restrict ourselves
too excessively to particular prejudices regarding physics as yet unknown
(though, of course, the notion of triviality does require the assumption
of a SUSY desert up to the unification scale).

Our calculations suggest that the universal
upper bound on the lightest CP-even Higgs mass is lifted from 145 GeV to
149 GeV, a mere 4 GeV difference. Table \ref{table} reveals that for
larger values of $m_t$ the shifts may be up to 23 GeV, but for small $m_t$
the squark effects are small, as expected. We emphasise that we have not
re-maximised the complete one loop corrected bound, but rather made the
assumption that the values of $h_t$ and $\lambda_{max}$ which generated the
bound of our previous analysis are not significantly modified by the
inclusion of a general spectrum of squark masses.

To conclude, we have supplemented our previous RG analysis of the Higgs
sector of the NMSSM with a calculation of general squark effects, and
determined how they modify the bound on the mass of the lightest CP-even
state. Our calculations indicate that the effects of a general squark
spectrum can be very significant for large top quark masses. One reason why,
for large top quark masses, squark effects
are important is that there exist finite, one--loop diagrams with vertices
containing the uncontrolled soft SUSY breaking parameter $A_t$. These
diagrams can give large contributions to the Higgs boson masses. We close
by indicating that in the NMSSM there are similar diagrams involving loops
of Higgs bosons, again with vertices containing soft SUSY breaking
parameters,
this time $A_{\lambda}$ and $A_k$. The existence of these diagrams is
directly due to the gauge singlet field $N$; they do not occur in the MSSM.
We see no reason in principle why these new diagrams should not give rise
to similarly large effects, except for obvious factors of 3 due to colour.
We are currently in the process of estimating these effects \cite{elliott2}.

\begin{center} {\bf Acknowledgements} \end{center}

\noindent S.F.K. is grateful to the S.E.R.C. for the support of an Advanced
Fellowship, and T.E. is grateful to the S.E.R.C. for the support of
a studentship.

\clearpage

\section*{Figure Caption}

\noindent {\bf Figure 1:} The bound on the lightest CP-even Higgs boson mass
in the NMSSM. The upper line is the bound $m_h$ including a general
spectrum of
squark masses. The lower line is the bound $m_h^0$ of our previous analysis,
from assuming degenerate squarks with a common mass $M_{SUSY}$.

\end{document}